\documentclass[review]{elsarticle}

\usepackage{hyperref}
\usepackage[font=small,labelfont=bf,tableposition=top]{caption}
\usepackage[caption=false]{subfig}% Support for small, `sub' figures and tables
\usepackage{amsmath, amssymb}
\usepackage{cleveref}
\usepackage{xcolor}

\journal{Special Issue on ''Modeling and forecasting of epidemic spreading''}

\crefformat{equation}{Eq. (#1)}
\crefformat{figure}{Fig. #1}
\crefmultiformat{figure}{Figs. #1}
{ and #2#1#3}{, #2#1#3}{ and #2#1#3}
\crefformat{table}{Table #1}
\crefformat{appendix}{#1}
\crefformat{section}{Section #1}

\begin{document}

\begin{frontmatter}

\title{A non-homogeneous Markov early epidemic growth dynamics model. Application to the SARS-CoV-2 pandemic}

%\title{Elsevier \LaTeX\ template\tnoteref{mytitlenote}}

\author[yo]{N\'estor Ruben Barraza\corref{mycorrespondingauthor}}
\fntext[myfootnote]{The author is also with the School of Engineering of the University of Buenos Aires.}
\cortext[mycorrespondingauthor]{Corresponding author}
\ead{nbarraza@untref.edu.ar}
\author{Gabriel Pena}
\author{Ver\'onica Moreno}
\address{Universidad Nacional de Tres de Febrero}




\begin{abstract}
This work introduces a new markovian stochastic model that can be described as a non-homogeneous Pure Birth process. We propose a functional form of birth rate that depends on the number of individuals in the population and on the elapsed time, allowing us to model a contagion effect. Thus, we model the early stages of an epidemic. The number of individuals then becomes the infectious cases and the birth rate becomes the incidence rate. We obtain this way a process that depends on two competitive phenomena, infection and immunization. Variations in those rates allow us to monitor how effective  the actions taken by government and health organizations are. From our model, three useful indicators for the epidemic evolution over time are obtained: the immunization rate, the infection/immunization ratio and the mean time between infections (MTBI). The proposed model allows either positive or negative concavities for the mean value curve, provided the infection/immunization ratio is either greater or less than one. We apply this model to the present SARS-CoV-2 pandemic still in its early growth stage in Latin American countries. As it is shown, the model accomplishes a good fit for the real number of both positive cases and deaths. We analyze the evolution of the three indicators for several countries and perform a comparative study between them. Important conclusions are obtained from this analysis.
\end{abstract}


\end{frontmatter}

\section{Introduction}

The coronavirus disease (COVID-19) transmitted by the SARS-CoV-2 appeared in China by the end of 2019. More than 17000000 of cases and 675000 deaths were reported since then. The lack of a vaccine and an effective treatment for this disease forced the governments to adopt lockdown actions in order to protect the population and to slow down the spread of the outbreak. Despite those actions, health systems in several countries resulted overwhelmed. Many mathematical models were developed in order to predict the behavior of the outbreak in several ways, most of them based on the well known SIR (Susceptible-Infectious-Recovered) models. We present in this work a different approach having the advantage of a rapid and clear interpretation of its parameters. We obtain this way a quite useful new indicator, the mean time between infections (MTBI).

Modeling contagion has attracted the interest of statisticians for decades. These models are useful to model the spread of contagious diseases in a population or plantation. A well-known example is the Polya urn model, where balls of different colors are extracted from an urn in such a way that when a given ball is extracted, not only is it put back but also a certain number of balls of the same color is added. The probability of getting a ball of that color in the next drawing is thus increased, modeling a contagious process. Contagion modeling has been already applied in many areas of engineering, see for example \cite{7789377,340476,784436}. The Polya stochastic process is obtained as the limit from the Polya urn model in a similar way as the Poisson process is obtained from a Bernoulli process.

As it happens with the homogeneous and non-homogeneous Poisson processes, the Polya stochastic process can be described by the Pure Birth equation (see chapter 17 of \cite{Feller1}). The main difference between these two types of Pure Birth processes is that, in the Polya process, the event rate is not just a function of time but also of the number of individuals in the population. There are also cases where the event rate depends only on the number of previous events but not on time, like the Yule process. However, since the probability of births rises due to an increase in the population while the probability of an individual birth remains constant, it does not describe a true contagion process.

The stochastic mean of the Polya process is a linear function of time, as will be shown. This characteristic makes the Polya process unsuitable for application in many real cases of disease growth dynamics and engineering. Hence, we propose a different model based on a Pure Birth process with an event rate that, like Polya's, depends on both the elapsed time and the number of previous events, but with a different functional form. We obtain this way a mean number of events that is a nonlinear function of time, with either an increasing or decreasing first derivative provided a given parameter is greater or less than one. 

An important application of our proposed model is the study of the spreading of diseases. We obtain this way a process controlled by two competitive phenomena, infection and immunization, i.e. two opposing forces. Hence, we get a model with two parameters: the transmission and immunization rates. The mean value function shows different behavior according to whether the ratio between infection and immunization rates is greater or lower than one. In the former case, its second derivative is positive and the mean value curve is convex, whereas on the latter case its second derivative is negative and the resulting curve is concave. The limit case of infection/immunization ratio $\gamma/\rho=1$ is also possible, resulting on a mean value function that grows linearly over time. The evolution of the parameters shows how well the pandemic is being controlled, either by lowering the infection rate via isolation and quarantine measures or by increasing the immunization rate with a vaccination campaign. We obtain this way three useful indicators: the infection/immunization ratio, the immunization rate and the mean time between infections (MTBI).

Our model is quite suitable to be applied to early epidemic growth dynamics, where exponential models fail (this matter is discussed in \cite{CHOWELL201666}). Since ours falls in the category of subexponential models, the results presented in this article support the idea developed in the cited reference. Another advantage of our approach is that, as a subset of the subexponential cases, we are able to model cases with linear and sublinear (concave) cumulative case incidence curves, as it will be shown.

As an interesting and quite important application, we apply our model to the recent SARS-CoV-2 pandemic as an alternative to the SIR models. Our theoretical values match real data, which allows us to assess the evolution of the oubtreak and to predict its trend a few days ahead. The parameters can be easily estimated through simple methods, such as Least Squares, applied either to positive cases or deaths. We show applications to data from several countries with good agreement. 

This paper is organized as follows: Motivation and related work are exposed in \cref{mr}, the proposed model is presented in \cref{promodel}, applications to the SARS-CoV-2 pandemic are developed in \cref{asr} and some discussions on the results we obtained are presented in \cref{dc}. The final conclusions are exposed in \cref{conc}.

\section{Motivation and Related work}\label{mr}

\subsection{Motivation}

Our main motivation is to obtain a model that describes an epidemic outbreak at its first stage, before it reaches the inflection point in the case incidence curve, which is useful to monitor how contagion is spreading out. This first stage corresponds to an $R$ naught greater than one. Our model is inspired in the Polya-Lundberg process, which comes from the contagious urn model as explained in the previous section. Since the mean value function of the Polya-Lundberg process is a linear function of time (see \cref{apMean}), we introduce a modification in the event rate in order to get a mean value function that grows subexponentially with either positive or negative concavity as we observe in the early epidemic growth curves usually reported.  

\subsection{Related work}

There exists a wide variety of models that can be used to describe the evolution of an epidemic. Main standard is held by the so called compartmental models, i.e., the family of SIR based models (SIR, SEIR, SIRS, etc. See for instance \cite{LiModelingDiseases,anderson1992infectious,ChenInfectiousDiseases,GETZ20189}). These models consist of a set of differential equations that take into account transitions between different compartments. The first classical SIR model was porposed by Kermack and McKendrick \cite{kermack}. 

Due to the recent COVID-19 coronavirus pandemic, most of the scientific community is dedicated to study its behavior, both by using SIR based models and introducing new ones. In \cite{ALI}, a stochastic term is introduced in the system of differential equations to simulate noise in the detection process. In \cite{CONTRERAS} the authors use an ARIMA based method to correct errors arising from the delay between the day the sample is taken and the day a diagnosis is made, which contaminate the time series. The impact of statistical assumptions within the parameter inference for the phenomenological models is analyzed in \cite{GANYANI2020110029}.

More recent approaches to the problem of assessing the behavior of an epidemic include computational simulation and machine learning. Agent-Based Models (ABM) perform simulations of a population made by a large number of individual agents who can take decisions based on some criteria. In \cite{SILVA} the authors propose a SEIR ABM which simulates the behavior of a society under different scenarios in order to forecast both the economic and epidemic impact of possible government measures. A comparison between ABM and classical compartmental models can be found in \cite{VanDykeParunak}. In \cite{RIBEIRO, DASILVA} some well-known machine learning and time series algorithms (like ARIMA and SVR) are used to learn from a present dataset and forecast the case incidence curve up to the following six days.
Combinations of SIR models embedded with some random components was proposed in very recent articles, see for instance \cite{SIMON20204252,CAO2020124628,liulevy}. A theoretical framework to model epidemics by a non homogeneous birth-and-death process was proposed in \cite{CLANCY1998233}.

An extensive analysis of epidemics at their early stage was performed in \cite{CHOWELL201666}. In this study, the authors propose a phenomenological model (see also \cite{VIBOUD201627,PELL201862}) and analyze the application of SIR based models with either exponential and subexponential behavior. Our approach is quite different. On one hand, our model is stochastic whereas theirs is deterministic. On the other hand, we obtain our contagion model as a special case of a very general and well-known probabilistic model, the Pure Birth process.
It is interesting to compare the differential equations the cited authors arrive with ours, which will be done in a future publication. Furthermore, and unlike them, we are also able to model cumulative case incidence curves that grow with negative concavity.

\section{The proposed model}\label{promodel}

The model we propose is based on a Pure Birth process. These processes describe the evolution of a population where individuals can only be born. We are then modeling the epidemic spread as births in a population, where every birth corresponds to a new infection case. Our novel approach consists of finding a proper incidence rate that describes the contagion phenomenon. We develop next the fundamentals of Pure Birth processes and our proposal.  

\subsection{Pure Birth processes}

We propose to get the probability of having $r$ infections in a given time $t$ from a Pure Birth stochastic differential equation. Pure Birth processes describe the behavior of a population where individuals can only be born and are not allowed to die. All the individuals are assumed to be identical. The probability of having $r$ individuals in the population at a given time $t$, $P_r(t)$, is given by the solution of the following differential equation, see \cite{Feller1,lawler2006introduction}:

\begin{align}
	\label{purebirth}
	P'_r(t) &= -\lambda_r(t) P_r(t) + \lambda_{r - 1}(t) P_{r-1}(t), \quad r\geq 1 ,\nonumber \\
	P'_0  (t)&=-\lambda_0(t)P_0(t).
\end{align}

Assuming we have $0$ individuals at time $0$, we impose the following initial conditions on \cref{purebirth}: $P_0(0)=1$ and $P_r(0)=0$ for $r\geq 1$. To see that $ P_r(t)$ is well defined see \cref{ap1}.

The process given by \cref{purebirth} is also markovian, where the number of individuals corresponds to the state of the system, $S_r(t)$. The only transitions allowed{\footnotemark} are $S_r(t) \rightarrow S_r(t + dt)$ and $S_r(t) \rightarrow S_{r + 1}(t + dt)$ with probabilities:

\begin{align}
	\label{probtrans}
	P(S_r(t) \rightarrow S_{r + 1}(t + dt)) &= \lambda_r(t) \; dt ,\nonumber \\
	P(S_r(t) \rightarrow S_r(t + dt)) &= 1 - \lambda_r(t) \; dt.
\end{align}

\footnotetext{Actually, other transitions have probabilities that are higher order infinitesimals.}

The dependence of $\lambda_r(t)$ on $t$ defines the type of process. If $\lambda_r(t)$ is a function of $t$ only or a constant, then the process is non-homogeneous or homogeneous respectively, with independent increments in both cases. If $\lambda_r(t)$ is also a function of $r$, the process has dependent increments.

From \cref{purebirth}, the probability of having no births in a certain time interval greater than $t - s$ given the population has $r$ individuals by the time $s$ is given by the well-known exponential waiting time:
\begin{equation}
	\label{nobirths}
	P({\rm no \; births \; in \; } T \ge t - s) = e^{- \int_s^t \lambda_r(t) dt} \;\;\;\;\; t \ge s.
\end{equation}

The mean number of infections in a given time is
\begin{equation}
	\label{mean-r}
	M(t) = \sum_r r \; P_r(t).
\end{equation}

Under certain conditions (see \cref{apMean} to verify those are satisfied by our model) we have the following differential equation for the mean value:

\begin{equation}
	\label{mean-de}
	M'(t) = \sum_{r = 0}^\infty \lambda_{r }(t) P_{r}(t).
\end{equation}

Depending on the proposed function for $\lambda_r(t)$, the mean number of failures may or may not be easily obtained.  

Our formulation allows us to calculate the mean time between infections (MTBI), this is, the mean time between births in a Pure Birth process. From \cref{nobirths}, we can predict the mean time between infections (MTBI) after $r$ infected individuals were detected by the time $s$ using a model with incidence rate $\lambda_r(t)$ as:
\begin{equation}
	\label{eq_mtbi}
	MTBI(s,r) = \frac{1}{Z} \; \int_s^\infty \; \lambda_r(t) \; t \; e^{- \int_s^t \lambda_r(t) dt} \;\; dt - s.
\end{equation}

Here, $Z$ is a normalizing constant to consider cases where the probability of having no infections in an infinite time interval is greater than zero and is given by \cref{znorm} (see \cref{ap-MTBI} for a full deduction):

\begin{equation}
	\label{znorm}
	Z = 1 - e^{- \int_s^\infty \lambda_r(t) dt}.
\end{equation}

\cref{eq_mtbi} gives the expected time to the next infection given that by the time $s$ the infected population consists on $r$ individuals, this is, the mean time between two consecutive detections. 

\begin{figure}[h!]
	\centering
	\includegraphics[width = 0.8\textwidth,page=27]{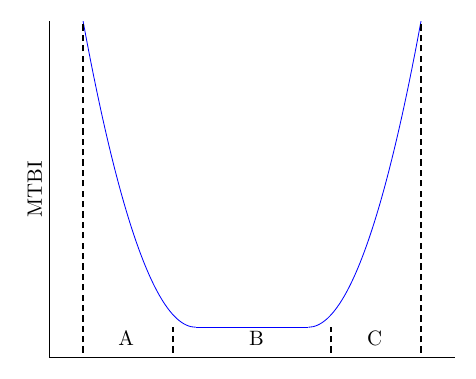}
	\caption{Expected shape of the MTBI indicator. Different portions show the following: A-The epidemic is spreading out, MTBI decreases since infections occur more often. B- Infections occur at a constant rate. C- MTBI increases since infections occur less often.}
	\label{mtbi-shape}
\end{figure}

This indicator is, in most cases, U-shaped (see \cref{mtbi-shape}). An exceptional case is also shown in \cref{uruguay_mtbi}, and will be discussed later). It is expected to decrease at first, showing the acceleration of the spread (infections occur more often). When the cumulative case incidence curve reaches the inflection point and the epidemic starts to mitigate, MTBI shows a flattening portion (infections occur at a constant rate). Then, due to the deceleration of the epidemic (infections occur less often) MTBI tends to get larger again. This curve resembles the bathtub curve, well known in Reliability Engineering, which is also obtained when this class of models is applied to reliability studies, see \cite{penabarrazaanand}. 

\subsection{Proposed incidence rate}

We propose a two-parameter incidence rate given by the following expression:
\begin{equation}
	\label{proposedfr}
	\lambda_r(t) = \rho \frac{1 + \frac{\gamma}{\rho} \; r}{1 + \rho \; t}.
\end{equation}

The parameters involved in \cref{proposedfr}, indicate how fast the outbreak is spreading. We can see that the $\gamma$ parameter is related to the strength of the contagion effect, since it is a factor of the number of infections, and the $\rho$ parameter is related to the outbreak mitigation, either by natural immunization or due to external measures. That is the way how these two parameters, once estimated, are useful to monitor the disease progress and the impact of actions taken by health institutions. These actions can affect either the $\gamma$ parameter via the population solation and quarantine, or the $\rho$ parameter by, for instance, vaccination campaigns. As we will show, the ratio between both rates is the exponent of time $t$ on the mean value function (\cref{mean-r-our}), and it determines the curve's concavity, provided it is greater or less than one. We can compare our proposed incidence rate with other formulations by rewriting \cref{proposedfr} as follows:

\begin{equation}
	\label{incidencerate}
	\lambda_r(t) = \frac{\rho}{1 + \rho \; t} + \frac{\gamma \; r}{1 + \rho \; t}.
\end{equation}
The numerator of the second term in \cref{incidencerate} can be compared with the numerator of the incidence rate analyzed in \cite{Mishra87} (Eq. 1: $\lambda(t) = \beta \; c \; \frac{I(t)}{N(t)}$, where $I(t)$ is the number of infections, $\beta$ the transmission probability per contact and $c$ the contact rate) by replacing $I(t)$ with $r$ and $\beta \; c$ with $\gamma$; our $\gamma$ parameter can be interpreted as the transmission rate in the same way. In that proposal, the $N(t)$ denominator corresponds to the total population, since the authors consider the per capita incidence rate. As an interesting remark, it can be seen that it also coincides with that of the Polya-Lundberg process. The first term in \cref{incidencerate} is equal to $\lambda_0(t)$, hence it can be considered as the initial incidence rate. On the other hand, following the usual behavior of $\rho$, this term rapidly vanishes with time.

Being the ratio $\frac{\gamma}{\rho}$ the coefficient of $r$ in the contagion rate (\cref{proposedfr}), it takes into account both the infection and immunization rates,  so it can be directly related to the well-known effective reproduction number $R_t$, an indicative that is often used to qualify the spreading and magnitude of an epidemic. The denominator in \cref{proposedfr} shows how the outbreak is getting weak as a function of time and is an indicator of how this Effective-reproduction-number-like parameter goes down. 

The evolution of these two parameters, the ratio $\frac{\gamma}{\rho}$ and $\rho$ as a function of time, is indicative of how well the epidemic is being controlled by actions taken from health institutions. We expect a strong decrease in $\frac{\gamma}{\rho}$, and in consequence, a strong increase in $\rho$. 

Our definition of the outbreak parameters deserves a detailed analysis. The well-known definition of the basic reproduction number $R_0$ assumes all the population is susceptible and no immunization is present. This would imply $\rho$ to be zero; in that case, \cref{proposedfr} reduces to
\begin{equation}
	\label{yule}
	\lambda_r(t) = \lambda \; r.
\end{equation}
which is the event rate of another classic stochastic model, called the Yule process. In that case, the event rate given by \cref{yule} grows linearly with the population size, which results in an exponential expression for the mean value function (see \cite{ross2008stochastic} for more details about the Yule process). This means that, if we allow our $\rho$ to eventually be zero, the Yule process becomes a special case of ours, hence exponential growth can also be modeled. However, as it can be seen in \cref{tab_infected}, even the worst case scenarios (very sharp incidence curves) differ from exponential growth, being this another empyrical confirmation of Chowell's thesis presented in \cite{CHOWELL201666}.

It should be remarked that our proposed rate (\cref{proposedfr}) is different from that of the Polya-Lundberg process: $\lambda_r(t) = \rho \frac{\frac{\gamma}{\rho} \; + r}{1 + \rho \; t}$, (see for example \cite{7789377}), which leads to a mean value function that depends linearly on time when replaced in \cref{mean-de}, as demonstrated in \cref{apMean}.

\subsection{The Case Incidence curve}

From the incidence rate (\cref{incidencerate}), we are able to get the exact solution of $P_r(t)$ (see \cref{ap1}) and therefore we can show that $M(t)$ is finite and \cref{mean-de} is valid (see \cref{apMean}). Solving \cref{mean-de} for our model we get the following expression for the mean (see \cref{apMean}):

\begin{equation}
\label{mean-r-our}
M(t) = \frac{\rho}{\gamma} ((1 + \rho \; t)^{\frac{\gamma}{\rho}} - 1).
\end{equation}

\begin{figure}
	\centering
	\subfloat[$\frac{\gamma}{\rho} > 1$]{\includegraphics[width = 2.5in,page=25]{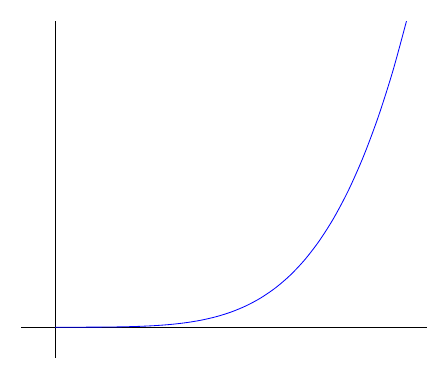}} 
	\subfloat[$\frac{\gamma}{\rho} < 1$]{\includegraphics[width = 2.5in,page=26]{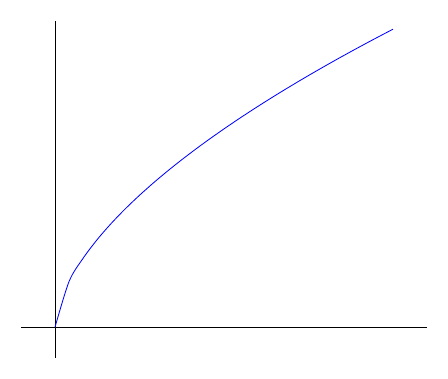}} %
	\caption{Different behaviors of \cref{mean-r-our} depending on values of $\frac{\gamma}{\rho}$.}
	\label{behavior}
\end{figure}

As seen from \cref{mean-r-our}, the mean value obtained from our model is a non-linear function of time with a positive or negative concavity whether $\frac{\gamma}{\rho}$ is greater or lower than one, as shown in \cref{behavior}. This expression is the functional form that describes the disease spreading at the early stage and fits well the COVID-19 data for both total infections and total deaths, as it will be shown next.

\subsection{The mean time between infections}

For our proposed model, it is straightforward to obtain an expression for the MTBI by inserting \cref{proposedfr} into \cref{eq_mtbi}. This leads to \cref{mtbi}:

\begin{equation}
	\label{mtbi}
	MTBI(r,t) = \frac{(1 + \rho \; t)}{\gamma \; r}.
\end{equation}

Replacing $r$ by the mean number of infections given by \cref{mean-r-our} we get \cref{mtbi-2}

\begin{equation}
	\label{mtbi-2}
	MTBI(t) = \frac{1}{\rho} \; \frac{(1 + \rho \; t)}{(1 + \rho \; t)^{\frac{\gamma}{\rho}} - 1},
\end{equation}

which is a useful conditional expression, as shown in \cref{ap-MTBI}.

\section{Applications to the SARS-CoV-2 disease}\label{asr}

In this section, we show the indicators obtained from several countries' reports. Data were taken up to early July 2020 from \url{https://ourworldindata.org/coronavirus-source-data}. For Asian and European countries, and also the USA, only data from the early epidemic stages were considered for analysis, this is, up to the inflection point. Real data and fitted curves are shown in \cref{fig_infected,fig_deaths}.

We fit \cref{mean-r-our} to the incidence and death curves of the recent CoV-2 pandemic in order to obtain the three mentioned indicators.

\subsection{Infection cases}

\begin{figure}[h!]
	\centering
	\subfloat[Argentina]{\label{incidence-argentina}\includegraphics[width = 1.8in,height=1.8in,page=4]{curvas.pdf}}
	\subfloat[Brazil]{\label{incidence-brasil}\includegraphics[width = 1.8in,height=1.8in,page=8]{curvas.pdf}}
	\subfloat[Chile]{\label{incidence-chile}\includegraphics[width = 1.8in,height=1.8in,page=9]{curvas.pdf}}\\
	\subfloat[China]{\label{incidence-china}\includegraphics[width = 1.8in,height=1.8in,page=5]{curvas.pdf}}
	\subfloat[Germany]{\label{incidence-germany}\includegraphics[width = 1.8in,height=1.8in,page=3]{curvas.pdf}}
	\subfloat[Italy]{\label{incidence-italy}\includegraphics[width = 1.8in,height=1.8in,page=1]{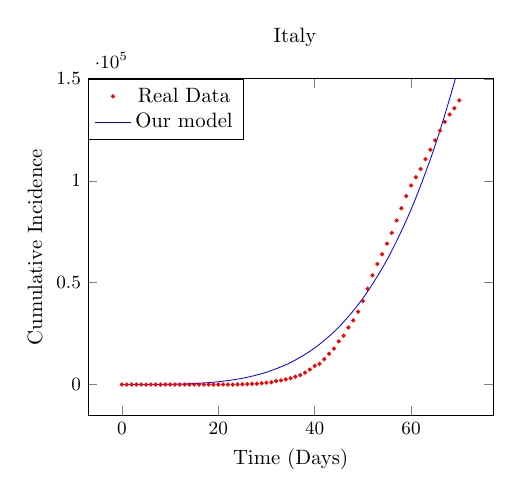}}\\ 
	\subfloat[Spain]{\label{incidence-spain}\includegraphics[width = 1.8in,height=1.8in,page=2]{curvas.pdf}}
	\subfloat[United Kingdom]{\label{incidence-uk}\includegraphics[width = 1.8in,height=1.8in,page=7]{curvas.pdf}}
	\subfloat[USA]{\label{incidence-usa}\includegraphics[width = 1.8in,height=1.8in,page=6]{curvas.pdf}}
	\caption{Cumulative case incidence.}
	\label{fig_infected}
\end{figure}

The case incidence curves are shown in \cref{fig_infected}, while \cref{tab_infected} shows the parameters. The coefficient of determination $R^2$ for measuring the goodnes of fit is also shown there. The model parameters were estimated by the Least Squares method, the resulting nonlinear equations being solved by the Levenberg-Marquardt algorithm.

\begin{table}[h!]
	\begin{center}
		\begin{tabular}[h]{lccc} 
			\hline\noalign{\smallskip}
			Country & $\rho$ & $\frac{\gamma}{\rho}$ & R$^2$\\
			\noalign{\smallskip}
			\hline
			\noalign{\smallskip}
			Argentina & 0.110403 & 4.72391 & 0.954157\\
			Brazil & 0.3629236 & 4.0432285 & 0.98094 \\
			Chile & 0.3139428 & 3.8070482 & 0.975592\\
			China & 0.310895 & 4.57184 & 0.944164\\
			Germany & 0.181481 & 5.0317 & 0.945185\\
			Italy & 0.573406 & 3.52942 & 0.932463\\
			Spain & 0.206731 & 4.93668 & 0.943004\\
			United Kingdom & 0.0274503 & 12.5826 & 0.975626\\
			USA & 0.0494053 & 9.53863 & 0.967517\\
			\hline
		\end{tabular}
		\caption{Parameter values and $R^2$ coefficient for the incidence curves.}
		\label{tab_infected}
	\end{center}
\end{table}

\begin{figure}[h!]
	\centering
	\subfloat[Incidence from Uruguay.]{\label{uruguay},\includegraphics[width = 0.49\textwidth,height=2.3in,page=17]{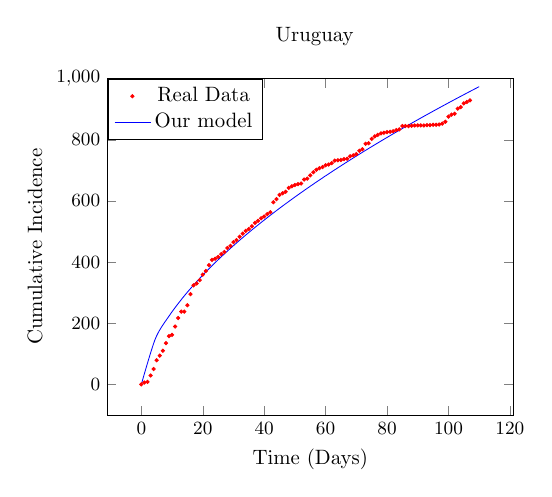}}
	\hfill
	\subfloat[MTBI from Uruguay.]{\label{uruguay_mtbi},\includegraphics[width = 0.49\textwidth,height=2.3in,page=28]{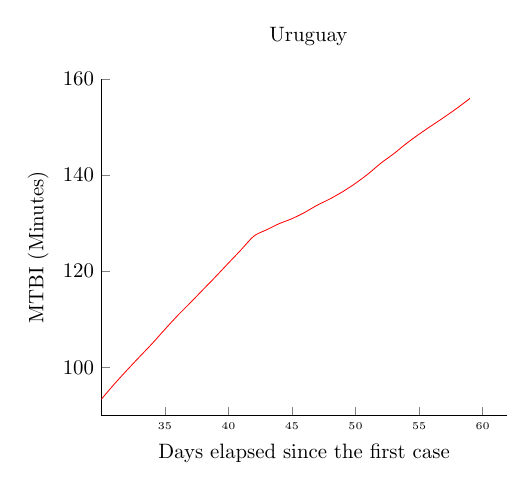}}
	\caption{The Uruguayan case.}
\end{figure}

In order to show that the proposed model fits well not just positive concavities as shown before, we consider the data from Uruguay, which presents a concave cumulative case incidence curve from the very beginning of the epidemic. Real data and the fitted curve are shown in \cref{uruguay}. Corresponding parameter values are $\rho = 459.363858194007$ and $\frac{\gamma}{\rho} = 0.586214262477143$. Another interesting case can be seen in \cref{south_korea_deaths}, which shows the number of deaths in South Korea. The curve is almost a straight line, and the corresponding $\frac{\gamma}{\rho} \approx 1$ as expected. Linear behavior is another limit case considered within our model. 

\subsection{Fatality cases}

\begin{figure}
	\centering
	\subfloat[Argentina]{\includegraphics[width = 1.8in,height=1.8in,page=5]{curvas-muertos.pdf}}
	\subfloat[Brazil]{\label{brasil_deaths}\includegraphics[width = 1.8in,height=1.8in,page=8]{curvas-muertos.pdf}}
	\subfloat[Chile]{\label{chile_deaths}\includegraphics[width = 1.8in,height=1.8in,page=9]{curvas-muertos.pdf}}\\
	\subfloat[Germany]{\includegraphics[width = 1.8in,height=1.8in,page=3]{curvas-muertos.pdf}}
	\subfloat[Italy]{\includegraphics[width = 1.8in, height=1.8in,page=1]{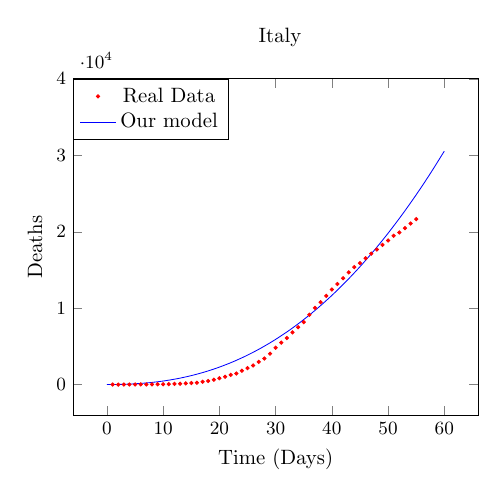}} 
	\subfloat[South Korea]{\label{south_korea_deaths}\includegraphics[width = 1.8in,height=1.8in,page=7]{curvas-muertos.pdf}}\\
	\subfloat[Spain]{\includegraphics[width = 1.8in,height=1.8in,page=2]{curvas-muertos.pdf}}
	\subfloat[United Kingdom]{\includegraphics[width = 1.8in,height=1.8in,page=6]{curvas-muertos.pdf}}
	\subfloat[USA]{\includegraphics[width = 1.8in,height=1.8in,page=4]{curvas-muertos.pdf}}
	\caption{Cumulative number of deaths.}
	\label{fig_deaths}
\end{figure}

Cumulative number of deaths curves are shown in \cref{fig_deaths}. The parameters and the coefficient of determination are shown in \cref{tab_deaths}.

\begin{table}
	\begin{center}
		\begin{tabular}[b]{lccc}
			\hline\noalign{\smallskip}
			Country & $\rho$ & $\frac{\gamma}{\rho}$ & R$^2$\\
			\noalign{\smallskip}
			\hline
			\noalign{\smallskip}
			Argentina &	0.166995 & 2.69049 & 0.962536\\
			Brazil & 1.26084 & 2.43167 & 0.9591606\\
			Chile & 0.07983 & 4.62764 & 0.9798674\\
			Germany	& 0.407652 & 3.36218 & 0.958036\\
			Italy & 1.75157 & 2.40184 & 0.843577\\
			South Korea	& 1.25081 & 1.35826 & 0.97169\\
			Spain & 2.17936 & 2.37205 & 0.916289\\
			UK & 0.258118 & 4.46389 & 0.960576\\
			USA	& 0.153287 & 5.74097 & 0.972502\\
			\hline
		\end{tabular}
		\caption{Parameter values and $R^2$ goodness of fit for the curve of deaths.}
		\label{tab_deaths}
	\end{center}
\end{table}

\subsection{Infection/immunization ratio over time}

In this section we show the evolution of the $\frac{\gamma}{\rho}$ parameter obtained from our model as a function of time, for the analyzed countries. In order to obtain the evolution of this parameter over time, we perform several estimation runs with successive subsets of the dataset and record each estimate.

This picture is indicative of how effective the measures taken by different countries have been, or how fast the outbreak is reaching the mitigation stage. As the transmission rate approaches the immunization rate, this parameter reaches the value of 1 as a limiting case. Curves are depicted in \cref{r0}.

\subsection{Immunization rate over time}

\begin{figure}
	\centering
	\subfloat[Infection/immunization ratio $\frac{\gamma}{\rho}$ as a function of time.]{\label{r0},\includegraphics[width = 3.5in,height = 3.2in,page=10]{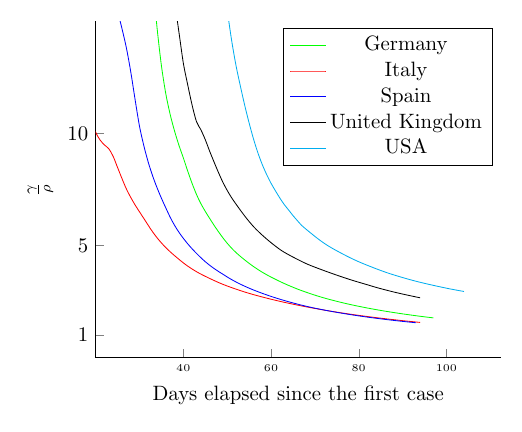}}\\
	\subfloat[Immunization rate $\rho$ as a function of time.]{\label{rt},\includegraphics[width = 3.5in,height = 3.2in,page=11]{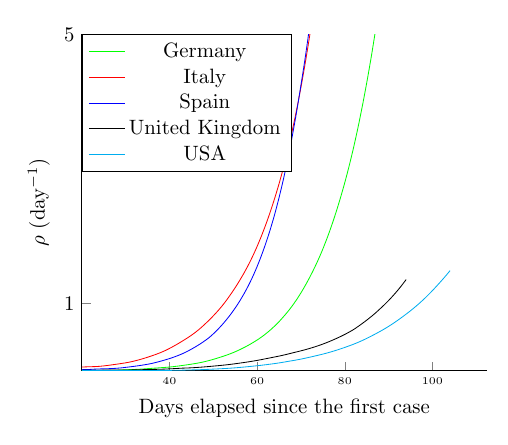}}
	\caption{Variation of the model parameters over time.}
\end{figure}

As previously discussed, the strength force that contains the outbreak is determined by the immunization rate $\rho$ parameter, and we expect it to increase due to actions taken by governments. Then, we analyze the evolution of this parameter in the same way as it was done for $\frac{\gamma}{\rho}$ in the previous section. Values of $\rho$ as a function of time are depicted in \cref{rt}.

\subsection{Latin American countries}

\begin{figure}
	\centering
	\subfloat[Argentina]{\includegraphics[width = 2.5in,height = 2.2in,page=21]{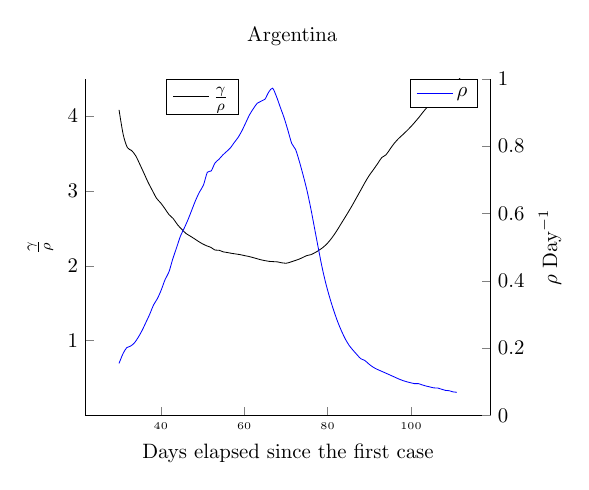}} 
	\subfloat[Brazil]{\includegraphics[width = 2.5in,height = 2.2in,page=20]{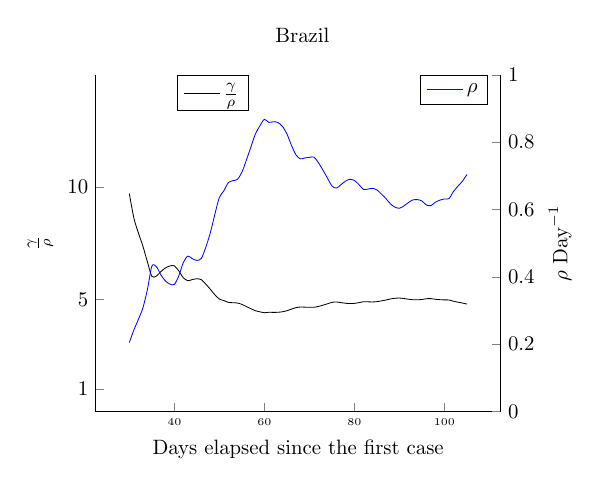}}\\
	\subfloat[Chile]{\includegraphics[width = 2.5in,height = 2.2in,page=22]{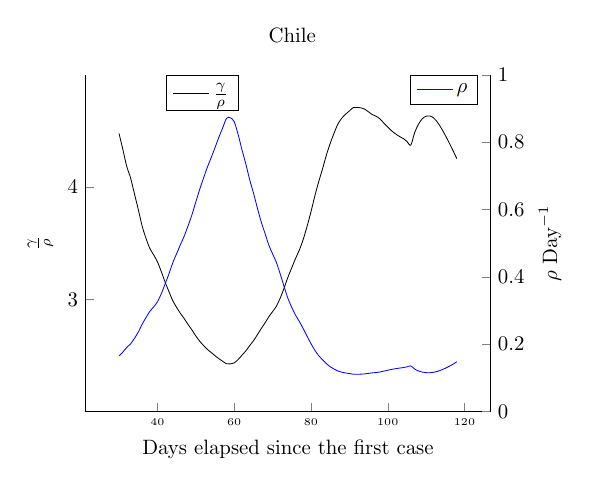}} 
	\subfloat[Peru]{\includegraphics[width = 2.5in,height = 2.2in,page=29]{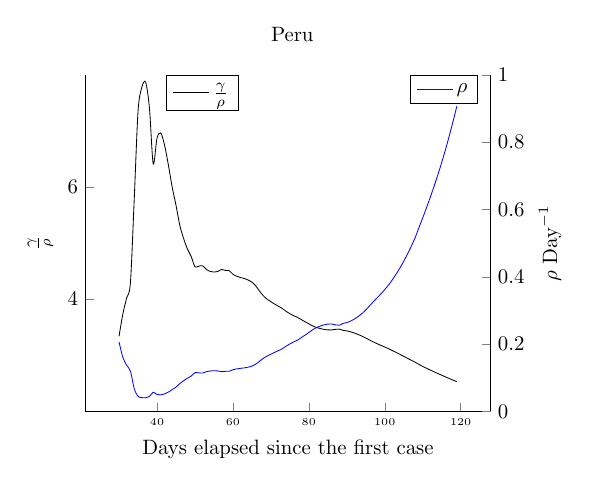}} 
	\caption{Parameters as a function of time.}
	\label{rt-latinos}
\end{figure}

It is interesting to analyze the current behavior of the parameters for Latin American countries, since they are still at the first stage of the pandemic and where, with the exception of Brazil, the lockdown was relaxed and turned strict again more than once. Curves are depicted in \cref{rt-latinos}.

\subsection{Mean time between infections}

\begin{figure}
	\subfloat[Argentina]{\includegraphics[width = 2.5in,height = 2.2in,page=12]{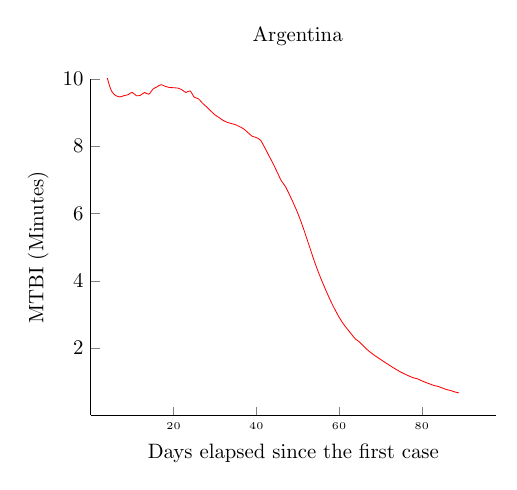}} 
	\subfloat[Brazil]{\includegraphics[width = 2.5in,height = 2.2in,page=24]{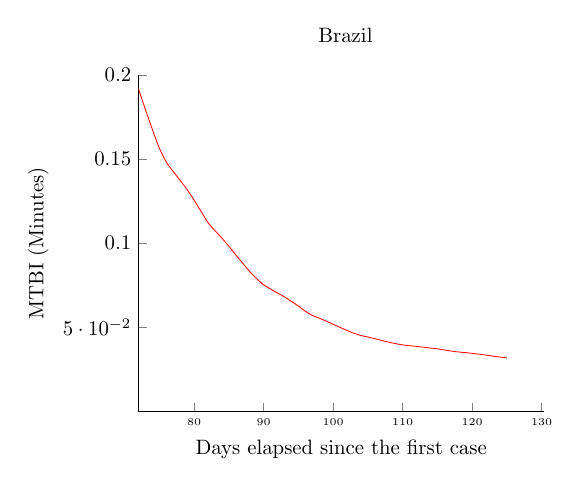}}\\
	\subfloat[Chile]{\includegraphics[width = 2.5in,height = 2.2in,page=23]{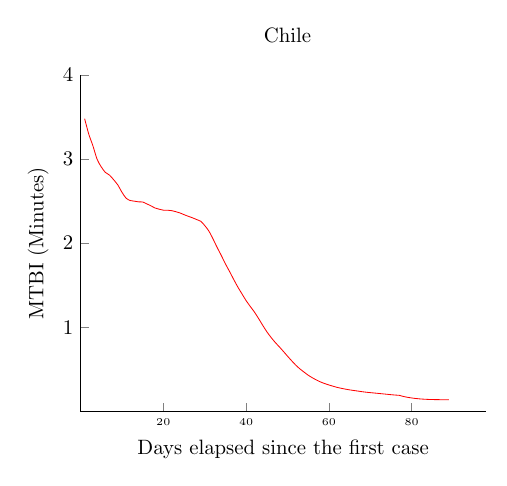}} 
	\subfloat[Italy]{\includegraphics[width = 2.5in,height = 2.2in,page=13]{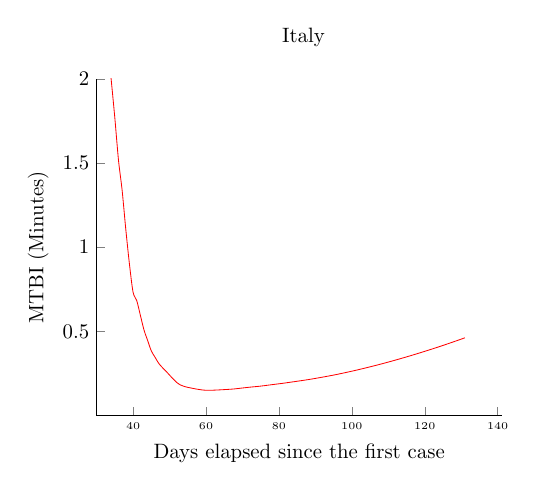}} 
	\caption{Mean time between infections (MTBI).}
	\label{mtbi-45}
\end{figure}

Values of the MTBI obtained from \cref{mtbi-2} for Argentina and Italy are shown in \cref{mtbi-45}. MTBI decreasing over time reflects the pandemic is still spreading out and the inflection point has not been reached yet. This indicator is useful in order to estimate the moment when the curve will start to flatten.

\begin{figure}[h!]
	\centering
	\includegraphics[width = 5in,page=14]{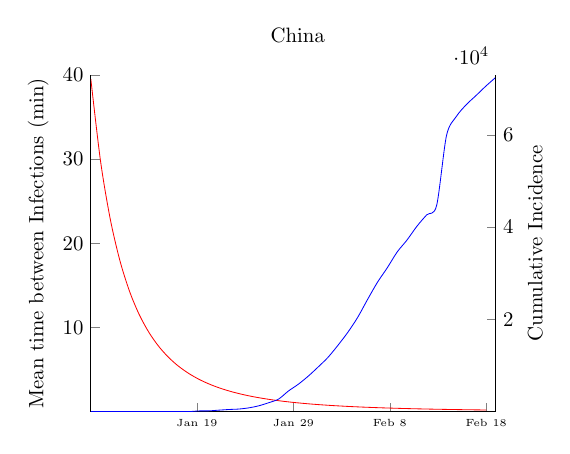} 
	\caption{Mean time between infections and cumulative cases incidence for China.}
	\label{mtbi-china}
\end{figure}

In \cref{mtbi-china} we depict the MTBI for China in order to see the flattening section of this indicator as the cumulative cases of incidence reaches its inflection point.

By the time we are writing this report (early July 2020), contrarily to Europe and the USA, Latin American countries have not yet reached the inflection point in the cumulative incidence curve, which can also be seen through the MTBI characteristics. This inflection point in the cumulative incidence curve corresponds to a minimum in the MTBI indicator, as seen in \cref{mtbi-45} for Italy. It is  interesting to observe the minimum values reached by several countries and compare those to the actual values of some Latin American ones. These are shown in \cref{tabla243}.

\begin{table}
	\begin{center}
		\begin{tabular}{l|c} 
			\hline\noalign{\smallskip}
			Country & Minimum MTBI [Minutes]\\
			\noalign{\smallskip}
			\hline
			\noalign{\smallskip}
			Argentina* & 0.7 \\
			Brazil* & 0.032 \\
			Chile & 0.14 \\
			China & 0.19 \\
			Germany & 0.14\\
			Italy & 0.14\\
			South Korea & 1.21 \\
			Spain & 0.10\\
			United Kingdom & 0.17\\
			USA & 0.024 \\
		\end{tabular}
	\end{center}
	\caption{MTBI. Minimum values reached. Countries marked with * have not reached a minimum by the time this article is being written; the present value is then indicated.}
	\label{tabla243}
\end{table}

\section{Discussion}\label{dc}

The well goodness of fit obtained for several countries shows that the spreading is governed by the same law. Although with different parameter values, the cumulative case incidence curve follows the same mathematical expression. Those parameters indicate the level of contagion and the effectiveness of the actions taken by governments. The model fits rather well the number of infected and deaths in the population previous to the inflection point, at the early pandemic growth stage, though its prediction beyond four or five days ahead is generally too pessimistic and overestimates the actual data.

The MTBI minimum values indicated in \cref{tabla243} show an interesting result. Countries from Europe and China present similar values, between 0.1 and 0.2 minutes, i.e. between 5 and 10 infections per minute. It must be remarked that this value does not correspond to the same vector or host: the Pure Birth process considers the population to be even. Hence, the MTBI value is the time elapsed between two consecutive infections, despite the agent that makes the infection. European countries and China have similar minimum values because those countries implemented quarantine actions rather late. Contrarily, Latin American countries like Argentina and Chile have implemented a lockdown at the beginning of the outbreak, reaching neither their minimum nor the values of the other regions yet. Brazil and USA deserve a more detailed analysis. Since these countries have not implemented a strict lockdown, the pandemic has strongly spreaded between their population and, in consecuence, the case incidence curve suddenly increased. This also results in a sudden decrease in the MTBI up to quite low values, between 26 and 41 infections per minute. We have seen then that our indicator shows similar values for countries that have implemented similar actions. An open source software application developed in R that implements our proposed model, and which was used to obtain the shown curves, is available at \cite{penagithub}.

\section{Conclusion}\label{conc} 

A new early epidemic growth dynamics model was proposed. It is a slight variation of the Polya stochastic process, which in itself is a limiting case of the Polya urn model for contagion. Like the Polya process, the proposed model is a Pure Birth process with a different functional form for the event rate. In opposition to the Polya process, the new proposal results in a stochastic mean that depends non-linearly over time. Applications to the SARS-CoV-2 pandemic still in the first outbreak stage in several countries were shown. Our model is based on two competitive processes: transmission and immunization. Those are taken into account through the corresponding rates as factors. These parameters are easily estimated by a curve fitting procedure. The mathematical expression obtained from the proposed model fits well the incidence and deaths curves for several countries, even those with a concave shape as it is the case of Uruguay. Our model provides a quick and easy way to obtain indicators that can be used to monitor and evaluate the impact of measures taken in order to control the outbreak: the infection/immunization ratio, the immunization rate and the mean time between infections. 

\section*{Acknowledgments}

This work was supported by Universidad Nacional de Tres de Febrero under grant 32/473 A. The authors want to thank the PAHO/WHO modeling group, chaired by Dra. Ana Rivi\'ere Cinnamond, and the TC Modeling group chaired by Dr. Vandemaele, Katelijn A.h., and specially to Dr. Oscar Mujica, MD Epidemiologist, Regional Advisor Epidemiology Pan American Health Organization, PAHO/WHO Washington, DC. USA., for important suggestions and discussions.

\appendix

\section{Pure Birth processes}\label{ap1}

It can be easily proved that \cref{purebirth} has a unique solution and is given by (see also \cite{sendova}):

\begin{align}
	\label{eq0}
	P_{r}(t) &= \int_{0}^t  \Big[e^{-\int_x^t \lambda_{r}(v)dv} \Big] \lambda_{r-1}(x)P_{r-1}(x) dx, \quad r \geq 1 ,\nonumber \\
	P_{0}(t) &= e^{-\int_0^t \lambda_{0}(v)dv}.
\end{align}
$P_r(t)$ is a probability mass function provided that $\sum_{r=0}^{\infty}P_r(t)=1$; this is shown by following the same steps as in section 4 of Chapter XVII from \cite{Feller1} (see also \cite{stroockmarkov}, section 4.3.2), where it is proved for the case $\lambda_r(t)$ depends on $r$ but not on $t$. The authors are working on a future publication where this property will be properly generalized to functions that depend on $t$ as well.

Recall our proposed functional form of $\lambda_r(t)$:
\begin{equation}
	\label{proposedfr2}
	\lambda_r(t) = \rho \frac{1 + \frac{\gamma}{\rho} \; r}{1 + \rho \; t}.
\end{equation}
Let the function $\mu_r(t)$ be:
\begin{equation}
	\label{eq_mu_r_t}
	\mu_r(t) = \int_0^t \lambda_r (v) dv = (1 + r \; \frac{\gamma}{\rho}) \; \ln{(1 + \rho \; t)}.
\end{equation}

It can be seen from \cref{eq_mu_r_t} that $\mu_r'(t)= \lambda_r(t)$ and $\mu_r(0)=0 \quad\forall r\geq 0$. We now define the auxiliary function
\begin{equation}
	\mu_{\Delta}(t) = \mu_{r + 1}(t) - \mu_r(t) = \frac{\gamma}{\rho} \; \ln{(1 + \rho \; t)},
\end{equation}

which does not depend on $r$. Note that $\mu_{\Delta}(0)=0$.

It is a fact that the following equality holds:
\begin{equation}
	\label{eq30}
	\int_0^t \; \lambda_{r}(x) \; e^{\mu_{\Delta}(x)} \; (e^{\mu_{\Delta}(x)} - 1)^r \; dx = \frac{(\frac{\rho}{\gamma} + r \; )}{r + 1} \;  (e^{\mu_{\Delta}(t)} - 1)^{r + 1}.
\end{equation}
The proof can be done by differentiating the right side of \cref{eq30} to show it's indeed a primitive of the left side's integrand. Since the functions are also equal on $t=0$, fundamental theorem of calculus yields that the equalty is valid for every $t$.

Considering the initial condition $P_0(0)=1$ (the beginning population is 0), the probability mass functions are given by:
\begin{equation}
	\label{eq31}
	P_{r}(t) = \frac{\Gamma(\frac{\rho}{\gamma} +r)}{r! \; \Gamma(\frac{\rho}{\gamma} )} \; e^{-(\mu_{r}(t) - \mu_0(0))} \; (e^{\mu_{\Delta}(t)} - 1)^r.
\end{equation}

This is demonstrated by induction over $r$, using \cref{eq31} as inductive hypothesis and $P_0(t)$ from \cref{eq0}. In fact,

\begin{align*}
	P_{r+1}(t) &=  \int_{0}^{t}e^{-(\mu_{r+1}(t)-\mu_{r+1}(x))}\lambda_{r}(x)P_{r}(x)dx \\
	&= \frac{\Gamma(\frac{\rho}{\gamma} +r)}{r! \; \Gamma(\frac{\rho}{\gamma} )} e^{-(\mu_{r+1}(t) - \mu_0(0))} \int_{0}^{t} \lambda_{r}(x) e^{\mu_{\Delta}(x)}(e^{\mu_{\Delta}(x)} - 1)^r(x)dx \\
	&= \frac{\Gamma(\frac{\rho}{\gamma} +r)}{r! \; \Gamma(\frac{\rho}{\gamma} )} e^{-(\mu_{r+1}(t) - \mu_0(0))} \frac{(\frac{\rho}{\gamma} + r \; )}{r + 1} \;  (e^{\mu_{\Delta}(t)} - 1)^{r + 1}.
\end{align*}

We can rewrite \cref{eq31} as:
\begin{equation}
\label{eq31_v2}
	P_{r}(t) = \frac{\Gamma(\frac{\rho}{\gamma} +r)}{r! \; \Gamma(\frac{\rho}{\gamma} )} \; e^{-(\mu_{r}(t) - \mu_0(0)-r\mu_{\Delta}(t))} \; (1 - e^{-\mu_{\Delta}(t)})^r.
\end{equation}
Using the fact hat $\mu_{r}(t)=\mu_0(t)+r\mu_{\Delta}(t)$ and replacing $\mu_{0}(t)$, $\mu_0(0)$ and $\mu_{\Delta}(t)$ by their expression yields:
\begin{equation}
	\label{eq32}
	P_{r}(t) = \frac{\Gamma(\frac{\rho}{\gamma} +r)}{r! \; \Gamma(\frac{\rho}{\gamma} )} \; \left(\frac{1}{1 + \rho \; t}\right) \; \left(1 - \left(\frac{1}{1 + \rho \; t}\right)^{\frac{\gamma}{\rho}}\right)^r.
\end{equation}

It should be noted that, since the expression \cref{eq0} is also valid for the case $\lambda_r(t) = \rho \frac{\frac{\gamma}{\rho} \; + r}{1 + \rho \; t}$, we can define $\mu_{r}(t)$ and $\mu_{\Delta}(t)$ so that the same properties are achieved. Therefore, following the same steps we get the expression of the pmfs for the Polya-Lundberg process. A similar procedure provides the pmfs for the Yule process. 

\section{Mean value function for Pure Birth processes}\label{apMean}

Multipliying \cref{purebirth} by $r$ and summing we obtain 
\begin{align*}
	\sum_{r=1}^{\infty} r \; P_r ' (t) &= \lim\limits_{R\rightarrow \infty} \sum_{r=1}^{R} r\; P_r ' (t) \\
	&= \lim\limits_{R\rightarrow \infty}-\sum_{r=1}^{R} r \;\lambda_{r}(t)\; P_r (t)+\sum_{r=0}^{R-1} (r+1) \;\lambda_{r}(t) \;P_{r} (t) \\
	&= \lim\limits_{R\rightarrow \infty}-R\; \lambda_{R}(t) \;P_R (t) +\sum_{r=0}^{R-1} \lambda_{r}(t) \;P_{r} (t).
\end{align*}

Since the condition
\begin{equation}
	\label{lim_r_lambda_r_P_r}
	\lim\limits_{r \to \infty} \; r \; \lambda_r(t) \; P_r(t) = 0.
\end{equation}
is attained for our model, we get
\begin{equation}
	\label{eq_mv_2}
	\sum_{r=1}^{\infty} r \; P_r ' (t) = \sum_{r = 0}^\infty \lambda_{r}(t) \;P_{r}(t).
\end{equation}
By using the expression of $P_r(t)$ for our model (\cref{eq32}), it is straightforward to see that $M(t)$ is finite using D'Alembert's criterion to show that the series  $\sum_{r=1}^{\infty} r P_r(t)$ converges. Therefore, the left side of \cref{eq_mv_2} is $M'(t)$. Then, we can recover the mean value function by solving the following differential equation:
\begin{equation}
	\label{Ap_mean_de}
	M'(t) = \sum_{r = 0}^\infty \lambda_{r}(t) \;P_{r}(t).
\end{equation} 

Replacing $\lambda_{r}(t)$ in \cref{Ap_mean_de} by \cref{proposedfr2}
\begin{align*}
	M'(t) &= \sum_{r = 0}^\infty \rho \frac{1 + \frac{\gamma}{\rho} \; r}{1 + \rho \; t} P_{r}(t)\\
	&= \frac{\rho}{1 + \rho \; t}\sum_{r = 0}^\infty P_{r}(t)+ \frac{\gamma}{1 + \rho \; t}\sum_{r = 0}^\infty r \; P_{r}(t)\\
	&= \frac{\rho}{1 + \rho \; t}+ \frac{\gamma}{1 + \rho \; t} \; M(t)\\
	&= \frac{\rho}{1 + \rho \; t}(1 + \frac{\gamma}{\rho} \; M(t)).
\end{align*}
we obtain \cref{eq-dif-mean-model}, which can be solved to retrieve our model's mean value function, \cref{mean-r-our}.
\begin{equation}
	\label{eq-dif-mean-model}
	M'(t)= \frac{\rho}{1 + \rho \; t}(1 + \frac{\gamma}{\rho} \; M(t)).
\end{equation}

The same procedure that leads to \cref{Ap_mean_de} is also valid for the Polya-Lundberg process. Replacing $\lambda_r(t)$ by its expression and solving the differential equation results in a linear mean value function. In the same way, with $\lambda_r(t) = \lambda \; r$, we get an exponential mean value function, which corresponds to the Yule process.

\section{Mean time between infections calculation for the contagion model}
\label{ap-MTBI}

We begin from the exponential waiting time (\cref{nobirths}), which we can rewrite as
\begin{equation}
	\label{nobirths-2}
	P(T_r \geq t | t_{r-1}) = \exp\left(-\int_{t_{r-1}}^{t} \lambda_{r-1}(u)du\right).
\end{equation}

This shows that the exponential waiting time is, in fact, the probability complement of $F(t)$, which is the distribution function of the (finite) time until the \textit{r}-th infections. We shall write this as

\begin{equation}
	\label{time_distribution}
	F(t, t < \infty | t_{r-1}) = 1 - \exp\left(-\int_{t_{r-1}}^{t} \lambda_{r-1}(u)du\right).
\end{equation} 

Moreover, the probability of having an \textit{infinite} time until the \textit{r}-th infections can also be deduced from \cref{nobirths-2}. We can then write its complement as

\begin{equation}
	\label{finite_time_probability}
	P(t_r < \infty | t_{r-1}) = 1 - P(t_r = \infty | t_{r-1}) = 1 - \exp\left(-\int_{t_{r-1}}^{\infty} \lambda_{r-1}(u)du\right).
\end{equation}

Dividing \cref{time_distribution} by \cref{finite_time_probability} we recover the  distribution function of the time variable $t$ conditioned on it being finite, which we can differentiate to obtain the pdf

\begin{equation}
\label{f_t}
		f(t|t_{r-1},t_{r}<\infty)=\frac{\lambda_{r-1}(t)\cdot\exp\left(-\int_{t_{r-1}}^{t_r} \lambda_{r-1}(u)du\right)}{1 - \exp\left(-\int_{t_{r-1}}^{\infty} \lambda_{r-1}(u)du\right)}.
\end{equation}

Note that all these probabilities are conditioned on the previous detected infection cases. Taking the conditional expectation of $t$ results in \cref{eq_mtbi}, and replacing \cref{proposedfr} therein yields \cref{mtbi}, where $r$ and $t$ denote $r-1$ and $t_{r-1}$ respectively. We shall write this more precisely as
\begin{equation}
	\label{mtbi-3}
	MTBI_r(r-1, t_{r-1})=\frac{1+\rho t_{r-1}}{\gamma(r-1)}.
\end{equation}

Under the assumption that exactly the \textit{expected} number of infections have been detected, we can replace $r-1$ in \cref{mtbi-3} to get \cref{mtbi-2}, which we rewrite as
\begin{equation}
	\label{mtbi-4}
	MTBI_r(t_r|r-1=\mu(t_{r-1})) = \frac{1}{\rho} \; \frac{(1 + \rho \; t_{r-1})}{(1 + \rho \; t_{r-1})^{\frac{\gamma}{\rho}} - 1}.
\end{equation}

This is clearly a conditional \textit{MTBI}. \cref{mtbi-4} is useful to predict future values of the MTBI, when the real $r-1$ is unknown.

\bibliography{mybibfile}

\end{document}